\documentclass[a4paper,fleqn,usenatbib]{mnras}

\usepackage[T1]{fontenc}
\usepackage{ae,aecompl}

\usepackage{graphicx}	
\usepackage{amsmath}	
\usepackage{amssymb}	

\newcommand{\mdot}{\dot{M}}
\newcommand{\msun}{\mathrm{M}_{\odot}}

\newcommand{\mdotu}{\rm M_\odot yr^{-1}}

\title[Faraday rotation in GRMHD simulations]{Faraday rotation in
  GRMHD simulations of the jet launching zone of M87}

\author[M. Mo{\'s}cibrodzka et al.]{
M. Mo{\'s}cibrodzka$^{1}$\thanks{E-mail: m.moscibrodzka@astro.ru.nl},
J. Dexter$^{2}$\thanks{E-mail: jdexter@mpe.mpg.de},
J. Davelaar$^{1}$,
H. Falcke$^{1}$
\\
$^{1}$Department of Astrophysics/IMAPP,Radboud University,
  P.O. Box 9010, 6500 GL Nijmegen, The Netherlands\\
$^{2}$Max Planck Institute for Extraterrestrial Physics,
Giessenbachstr. 1, 85748 Garching, Germany
}

\date{Accepted XXX. Received YYY; in original form ZZZ}

\pubyear{2016}

\begin{document}
\label{firstpage}
\pagerange{\pageref{firstpage}--\pageref{lastpage}}
\maketitle

\begin{abstract}
Non-VLBI measurements of Faraday rotation at mm wavelengths have been used to
constrain mass accretion rates ($\mdot$) onto supermassive black holes in the
centre of the Milky Way and in the centre of M87.  We constructed general
relativistic magnetohydrodynamics models for these sources that qualitatively
well describe their spectra and radio/mm images invoking a coupled jet-disk
system. Using general relativistic polarized radiative transfer, we now also
model the observed mm rotation measure (RM) of M87. The models are tied to the
observed radio flux, however, electron temperature and accretion rate are
degenerate parameters and are allowed to vary. For the inferred low viewing
angles of the M87 jet, the RM is low even as the black hole $\mdot$
increases by a factor of $\simeq100$. In jet-dominated models, the observed
linear polarization is produced in the forward-jet, while the dense accretion
disk depolarizes the bulk of the near-horizon scale emission which originates
in the counter-jet. In the jet-dominated models, with increasing $\mdot$ and
increasing Faraday optical depth one is progressively sensitive only to
polarized emission in the forward-jet, keeping the measured RM relatively
constant. The jet-dominated model reproduces a low net-polarization of
$\simeq1$ per cent and RMs in agreement with observed values due to Faraday
depolarization, however, with $\mdot$ much larger than the previously
inferred limit of $9\times10^{-4}\,\mdotu$. All jet-dominated models produce
much higher RMs for inclination angles $i\gtrsim30^\circ$, where the
line-of-sight passes through the accretion flow, thereby providing independent
constraints on the viewing geometry of the M87 jet.
\end{abstract}




\section{Introduction}

Synchrotron radiation is intrinsically linearly polarized. If the wave passes
through a magnetized plasma (a Faraday screen) the plane of
polarization rotates. The degree of rotation is wavelength-dependent, with an
observable rotation measure (RM) given by,
\begin{equation}\label{eq:far1}
RM = \frac{\chi(\lambda_1)-\chi(\lambda_2)}{\lambda_1^2-\lambda_2^2}
\end{equation}
where $\chi$ is the position angle of the polarization plane and $\lambda_1$
and $\lambda_2$ are two closely spaced observing wavelengths.
The Faraday rotation can be also expressed in terms of the Faraday screen's 
density and magnetic field strength integrated along the line of sight:
\begin{equation}\label{eq:far2}
RM = 10^{4} \frac{e^3}{2 \pi m_{\rm e}^2 c^4} \int f_{\rm
  rel} n_e B_{||} dl \, \, \, {\rm [rad \, m^{-2}]}
\end{equation}
where $n_{\rm e}$ and $B_{\rm ||}$ are the electron density (${\rm cm^{-3}}$)
and magnetic field component projected onto the line of sight (Gauss),
respectively, and $dl$ is the line element (cm) \citep{gardner:1966}.  For a
relativistically cold plasma $f_{\rm rel}=1$, where the plasma electrons are
relativistically hot ($\Theta_{\rm e}\equiv k_B T_e/m_e c^2 >1$), the
correction term becomes $f_{\rm rel} =\log(\Theta_{\rm e})/2\Theta_{\rm e}^2$
and the rotation is suppressed (e.g., \citealt{jones:1977}).  Notice that
Eq.~\ref{eq:far2} is valid only if $\chi(\lambda)\sim\lambda^2$, which
is satisfied for a Faraday optical depth $\tau_{\rm FR} = \int dl \rho_V
\ll 1$, where $\rho_V$ is the Faraday rotation coefficient.

The effect of Faraday rotation is commonly used to probe the plasma and
magnetic field strengths and directions in various astronomical objects, e.g.,
in Active Galactic Nuclei (AGN) jets \citep{anderson:2015}, in jet radio lobes
(e.g., \citealt{feain:2009}), in galaxy clusters \citep{feretti:2012} and in
ionized gas of the interstellar medium \citep{haverkorn:2014}.

There are a few measurements of RM near supermassive black holes. In
Sgr~A*, the Galactic center black hole, the measured $RM = -5 \times
10^5 {\rm rad \, m^{-2}}$ (\citealt{bower:2003},
\citealt{marrone:2007}) and $RM = -7 \times 10^4 {\rm rad \, m^{-2}}$
is found at the Bondi scale a few arcseconds away \citep{eatough:2013}. 
Other black holes with RM detections include 3C 84
with $RM=8 \times10^5 \,{\rm rad \, m^{-2}}$ \citep{plambeck:2014},
the core of M87 (hereafter M87*) with $RM<2 \times10^5 \,{\rm rad \, m^{-2}}$
\citep{kuo:2014}, and PKS 1830-211 with $RM \approx 10^7 \,{\rm rad \,
  m^{-2}}$ (\citealt{martividal:2015}, where the observation possibly
probes plasma 0.01 parsec away from the black hole).

All these RM are measured at millimetre wavelengths with non-VLBI
observations, so that the sources are not resolved. In Sgr~A*, M87*,
and 3C 84, the measured RM was then used to constrain the mass
accretion rate onto the central object (e.g., \citealt{bower:1999}).
The usual procedure to interpret the observed RM is
to integrate Eq.\ref{eq:far2} in the radial direction with a power-law
profile for $n_e$ and $B$ (usually from equipartition condition) from
a semi-analytical advection dominated accretion flow model (ADAF,
e.g., \citealt{narayan:1998}). The accretion rate onto the black hole
is then provided by the ADAF model \citep{marrone:2006}. Based on this
model, the measured RM from Sgr A* of $-5.6 \pm 0.7 \times 10^{5} \rm
rad \rm m^{-2}$ was used to estimate $\mdot = 5 \times 10^{-9} -
2\times 10^{-7} \, \mdotu$ where the lower and upper limits are for
different slopes of the electron density radial profiles which depend
on the presence and properties of an outflow assumed to be produced by
accretion process \citep{bower:2003,marrone:2006,marrone:2007}. In a
similar fashion, the upper limit for the M87* of $|RM| < 7.5
\times 10^5 \rm rad \rm m^{-2}$ have led to an accretion rate
constraint of $\mdot < 9 \times 10^{-4} \, \mdotu$
(\citealt{kuo:2014}, \citealt{li:2016}).

These accretion rate constraints assume that the polarized emission is
produced close to the black hole and then is Faraday rotated in the
extended accretion flow. In Sgr~A*, the compact emission region
\citep{doelemanetal2008}, observed scaling of $\chi \sim \lambda^2$,
and $\simeq 10$ per cent RM variability \citep{marrone:2007} support
this scenario. In other sources, especially M87*, the emission near the
black hole may not be produced by an accretion flow such as an ADAF,
but, e.g., by a jet, so that the existing measurements would be
probing the plasma flowing out from the central region.  One also has
to keep in mind that in case of black holes their accretion flow is
simultaneously the source of synchrotron radiation and the Faraday
screen (e.g., \citealt{beckert:2002}).  The polarization plane position
angle $\chi$ does not have to be constant with $\lambda^2$ as if it
was if the polarized source is behind an 'external' Faraday
screen. This is because the accretion flow has a complex structure
where self-absorption and depolarization of radiation take place.
Finally, the $\chi$ may be additionally alternated by interstellar
medium in the host galaxy. All above suggest that observed $\chi(\lambda)$
should be interpreted carefully.

In this work, we present polarized emission models based on combining an ADAF-like
model realized in three-dimensional general relativistic magnetohydrodynamics
(3-D GRMHD) simulations with polarized radiative transport ray-tracing 
calculations near a Kerr black hole.  The radiative transfer model predicts the
synchrotron maps of the plasma appearance near the black hole in all Stokes
parameters (I,Q,U,V) at various wavelengths.  We can therefore calculate what is
the change of $\chi\equiv arg(Q+iU)/2$ as a function of wavelength and compute
the model RM using Eq.~\ref{eq:far1}, without using approximate formulas, such
as Eq.~\ref{eq:far2}.  Contrary to previous RM modeling, in the present
RM model all relativistic effects will be included: the observer viewing angle
effects and the magnetic field geometry are naturally taken into account and the
mass accretion rate onto the black hole is self-consistently calculated within
the GRMHD simulation.

Past studies of polarized radiative transfer through GRMHD simulations (e.g.,
\citealt{broderick:2010}, \citealt{romans:2012}, \citealt{gold:2016},
\citealt{dexter:2016}) mostly focused on models of linear polarization $LP
\equiv \sqrt{Q^2+U^2}$ and circular polarization ($V$) from the Galactic
center black hole Sgr~A*. Only \citet{broderick:2010} shows RM maps based on
GRMHD simulations of the M87 jet but these are done at low frequencies (15 GHz) for
solutions extrapolated to large distances from the black hole.  To our
knowledge there are no GRMHD polarized images of the M87* at millimetre
wavelengths. Hence our study focuses on 3-D GRMHD models that were applied to the
M87* by \citet{moscibrodzka:2016} (hereafter M16), where we studied only
unpolarized intensity of radiation at various wavelengths.

Here we show that the low net linear polarization and RM of M87*
\citep{kuo:2014} are in good agreement with the expectations from those
models that best match the observed properties of the radio jet. The
total intensity submm-wave image arises from the
\emph{counter-jet}, due to strong light bending effects. This emission is
depolarized by Faraday rotation in the dense, cold accretion disk. As a
result, the net polarization mostly originates in the forward jet, which
produces little flux, leading to a low total net polarization fraction and RM values
consistent with the observed upper limits. The accretion rate in the models
can be much larger than found from the semi-analytical ADAF RM modeling, since
the RM originates in the forward jet rather than in the accretion flow itself.

The paper is organized as follows. In Sect.~\ref{sec:model} we briefly
describe the GRMHD simulation. The radiative transfer models are
described in Sect.~\ref{sec:model_rad}. We present our results in
Sect.~\ref{sec:results} and summarize them in Sect.~\ref{sec:dis}.

\section{Model of accretion flow and jet}\label{sec:model}

The details of the GRMHD model scaled to M87* are described in M16.
The simulations follow the evolution of a hot accretion accretion flow in the equatorial
plane of a spinning, supermassive black hole. The black hole angular
  velocity is $a_*\approx0.94$. 
The accretion disk and the black
hole are driving magnetized jets that are roughly aligned with the spin axis
of the black hole. The model studied
here is refered as to SANE in the current jargon (which stand for Standard
And Normal Evolution), having low-power jets. 
The models extend in radius from the black hole event
horizon to 240 $R_g$. 

The GRMHD simulations provide only the fluid pressure, which is dominated by
the protons. In a perfect fluid, the pressure in a grid zone gives a proton
temperature. For underluminous accretion flows, protons and electrons are not
necessarily well coupled. We have to assume an electron temperature as they
are not self-consistently computed in the GRMHD simulations, yet they are
essential in calculating synchrotron emission. The electron temperature is
parameterized by a coupling ratio, $T_{\rm p}/T_{\rm e}$ , between proton and
electron temperature.  We study a class of models in which both the
proton-to-electron temperature ratio and the mass accretion rate onto the black
hole are allowed to vary. For each temperature ratio we adjust the mass
accretion rate onto the black hole (which is achieved by multiplying the
plasma density by a constant number) so that the total flux emitted at mm
wavelengths is equal to 1 Jansky, consistent with mm observations of the source
(\citealt{doeleman:2012}, \citealt{kuo:2014}).

In addition, $T_{\rm p}/T_{\rm e}$ depends on plasma magnetization. 
Following M16 we assume that the 
electron temperature is 
\begin{equation}
\frac{T_{\rm{p}}}{T_{\rm{e}}}= R_{\rm{high}} \frac{\beta^2}{(1+\beta^2)} +
R_{\rm{low}} \frac{1}{(1+\beta^2)}
\end{equation}
where plasma
$\beta=P_{\rm{gas}}/P_{\rm{mag}}$ and $P_{\rm{mag}}= B^2/2$.  $R_{\rm{high}}$
and $R_{\rm{low}}$ are temperature ratios that parametrize the
electron-to-proton coupling in the weakly magnetized regions (disk, high
$\beta$ regions) and strongly magnetized regions (jet, low $\beta$ regions),
respectively. We assume constant $R_{\rm low}=1$ and we vary $R_{\rm high}$.  Since
the nature of the mm-wave emission in the M87* is unknown, our goal is to smoothly
transition from the disk emission dominated models (in which the electrons are
strongly coupled to protons everywhere) to the jet emission dominated models
(in which the electrons are strongly coupled to proton in the jet region and
weakly coupled to protons in the disk zone).

In this paper we consider models with $R_{\rm low}=1$ and
$R_{\rm high}=1,5,10,20,40,100$. This is the same as the models studied in M16, for which the
corresponding accretion rates are $\dot{M}=1\times10^{-4}, 4\times10^{-4},
1\times10^{-3}, 3\times10^{-3}, 5\times10^{-3}, 9\times 10^{-3} \,\mdotu$,
respectively.  The models are listed in Table~\ref{tab:errors}.

In all models, the observer viewing angle is fixed to $i=20\degr$ from the
black hole spin axis consistent with observations of the large scale jet
\citep{mertens:2016}.  The 1~mm emission maps based on unpolarized radiation transfer are
presented in our previous work.  Notice that model RH100 is consistent with
the broadband emission observed in M87* and the images of this model
resemble the VLBI observations of the source at 7 and 3.5mm.  RH100 is
therefore currently our best-bet model for the M87*.

\section{Model of polarized radiation}\label{sec:model_rad}

Polarized synchrotron emission is computed with the fully general relativistic
polarized radiative transfer ray-tracing scheme {\tt
  grtrans}\footnote{http://www.github.com/jadexter/grtrans}
\citep{dexter:2009,dexter:2016}. This public code has been extensively tested
which makes it the current standard of the general relativistic polarized
radiative transfer models. The code includes approximate synchrotron emission
and absorption coefficients in all Stokes parameters, as well as Faraday
rotation and conversion \citep{shcherbakov:2008} suitable for thermal
plasmas as modeled here.

The result of the calculation are the intensity maps overplotted with polarization angle
$\chi$ at two closely spaced frequencies around 230 GHz (220, 240GHz) as a function of
coordinates on the sky. We use these $\chi$-maps to calculate maps of RM. The maps
show the inner 26 $GMc^{-2}$ in the plane of the black hole where the most of
emission at this wavelength comes from. We also integrate the maps with Stokes parameters 
to simulate non-VLBI observables that are currently available. 

Faraday rotation depends on the plasma Faraday optical depth defined here as
$\tau_{\rm FR}= \int \rho_V dl$, where $\rho_V = 4 \pi e^3/m_e^2 c^2 n_e B f /
\nu^2$ is the plasma rotativity.
Formally $f$ is a function of electron temperature $\Theta_e=k_B T_e/m_e c^2$,
$\nu$, and $B$ (see Eq. B16 in Appendix B2 in \citealt{dexter:2016}) but
approximately $f \approx \Theta_e^{-2}$.  Varying $R_{\rm high}$ from 1 to 100
corresponds to increasing $f$ by a factor of $10^4$. In our models the plasma
density scales with the accretion rate and $n_e \propto \dot{M}$ and the magnetic
field strength scales as $B \propto \dot{M}^{1/2}$ ($\beta = \rm const.$).  In
all models, $\nu=230$ GHz and $dl \approx 100 R_g=100 GM_{\rm BH}/c^2 =
9\times10^{16}$ cm are fixed. Hence, $\tau_{\rm FR} \propto
\dot{M}^{3/2}/T^2$ or $\dot{M}^{3/2}
R_{\rm high}^2$.  In model RH1, $n_e=10^4 \,{\rm cm^{-3}}$,
$B=0.1$ Gauss, and $\Theta_e=50$ so the expected model Faraday depth is
about $\tau_{\rm FR} \approx 1$. But in model RH100, the expected $\tau_{\rm
  FR} \approx 10^6$.  In the models considered here, the Faraday optical depth
changes from less than unity to extremely high values. Since the polarization vector
$\chi$ oscillates every $\tau_{\rm FR} = 2\pi$, this means that in extreme
cases $\chi$ will rotate by $\sim 10^6 \times 2\pi$ over small
distances.

The radiative transfer equations become stiff when
$\Delta \tau > 1$ to any transfer coefficient between sampled geodesic points,
and we find that an integrator for stiff equations \citep[ODEPACK integrator
  LSODA,][]{odepack} fails when $\Delta \tau \gtrsim 100$, a condition often
reached in the high accretion rate models even when using an enormous number of points on each geodesic, $n =
25600$. The analytic solutions for Faraday rotation are oscillatory in
(Q,U,V). Failing to converge causes the amplitude of the oscillations to grow,
which can produce enormous positive or negative intensities, and/or
polarization fractions much larger than 100 per cent. \citet{romans:2012} got around
this problem in low frequency models of Sgr A* by integrating ``spherical''
polarized Stokes parameters rather than (Q,U,V). In the spherical Stokes
parameters, the amplitude and phase of the oscillatory Faraday solutions are
integrated separately. Rapid variations of Faraday depth between zones
then lead to convergence difficulties in the phase, but do not cause failures
in the integration. The form of the polarized radiative transfer equations in
spherical Stokes parameters is given in Appendix \ref{sphstokes},
which we have implemented in \textsc{grtrans}.

A reliable RM solution of the radiative transfer equations
has to be accurate to $\ll 1$ per cent to not miss entire rotations of the
polarization vector. The radiative transfer solutions are converged down to
1 per cent as demonstrated in \citet{dexter:2016} but in case of problems with
moderate Faraday depths ($\tau_{\rm FR} \le 100$).  In the limit of
very large $\tau_{\rm FR}$ the polarization angle and degree do not
converge with the number of points on a geodesic. We instead calculate
mean values and error bars averaged over a number of runs.

\section{Results}\label{sec:results}

\subsection{Convergence of radiative transfer solutions}
\begin{figure}
\begin{center}
\includegraphics[width=0.5\textwidth,angle=0]{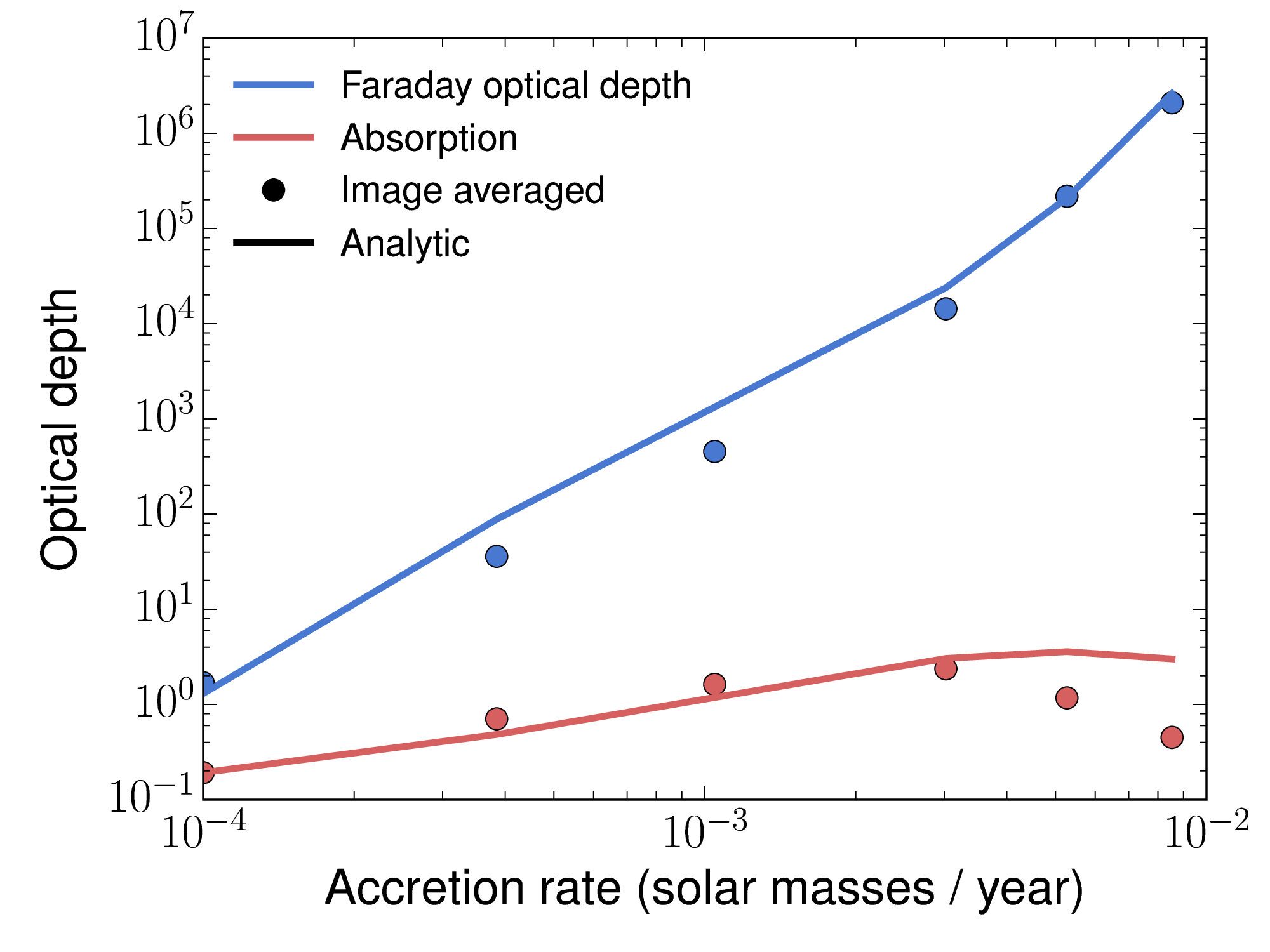}
\caption{Absorption and Faraday rotation optical depths as a function of
  accretion rate measured from intensity-weighted averages of all rays in the
  images (points) and estimated using $\tau_{\rm FR} \propto \dot{M}^{3/2} /
  T_{\rm disk}^2$ and $\tau_{\rm abs} \propto \dot{M}^{3/2} T$ 
  (solid lines), 
  scaled to the lowest $\dot{M}$ result. Models with lower disk temperature require higher
  accretion rates (M16), leading to rapidly increasing
  Faraday optical depth with accretion rate for these models while the absorption
  stays roughly constant.
}
\label{fig:tau}
\end{center}
\end{figure}

Figure~\ref{fig:tau} shows the expected vs. measured absorption and Faraday rotation
optical depth along the line of sight. The ``analytic'' curve is from the model $\dot{M}$
and $T_{disk}$, relative to the RH1 model, e.g. $\tau_{\rm FR} \propto
\dot{M}^{3/2} / T^2$, $\tau_{\rm abs} \propto \dot{M}^{3/2} T$, where the scaling
for absorption optical depth with $\dot{M}$ could also be $\dot{M}^2$. The
measured curves are from intensity-weighted averages along rays, which are
then also intensity-weighted across the images.  This shows how the Faraday
rotation optical depth grows dramatically with lowering disk temperature, and
is $\gg 1$ in all cases but RH1. The absorption on the other hand gets
important in the lower $T_{disk}$ models, but stays much smaller than the
Faraday optical depth.

It is necessary to demonstrate that the radiative transfer calculations
are fully converged in sense that numerical errors do not have a significant
contribution to the final result. To calculate the RM correctly, we need to have
maps converged in all Stokes parameters at nearby frequencies of interest.  We
check how well the solutions agree when simulating images with increasing
number of points along the geodesics, $n$ (logarithmically spaced within 50M
from the black hole).  In particular we consider two models with $n=32000$ and
$n=50000$ and calculate relative difference between them.

\begin{table*}
\centering
\caption{Convergence of radiative transfer based on models at $\nu_1=$230 GHz with n=32000 and
  n=50000. Image resolution in all models is $256^2$ pixels.
Columns 1 and 2 show the model ID and model mass accretion rate.
Columns 3 through 6 shows the difference between image integrated values in two runs with different number of integration points along geodesics lines. Column 7 shows the change of the polarization 
angle between  $\nu_1=$230 and $\nu_2=$240 GHz. 
Column 8 shows mean Faraday rotation between the two frequencies with errorbars from runs with n=8000, 16000, 32000, 50000  of integration points along the geodesics. 
Columns 9  and 10 show the total linear and circular polarization degrees of the model.}\label{tab:errors}
\begin{tabular}{cccccccccc}
\hline
ID  &$\mdot$ [$\frac{M_\odot}{yr}$] & $\frac{\Delta I_{tot}}{I_{tot}}$ &
$\frac{\Delta LP_{tot}}{LP_{tot}}$ & $\frac{\Delta V_{tot}}{V_{tot}}$
&$\Delta\chi_{tot}$ [rad]& $\chi_{tot,\nu_1}-\chi_{tot,\nu_2}$[rad] &
$\frac{\rm RM}{10^5}$ [${\rm \frac{rad}{m^{2}}}$ ] & $\frac{LP_{tot}}{I_{tot}}$ [\%] &$\frac{CP_{tot}}{I_{tot}}$ [\%] \\
(1) &(2) &(3) &(4) &(5) &(6) &(7) &(8) &(9)&(10)\\
\hline
RH1 & $1\times10^{-4}$ & $1.4\times 10^{-3}$ & $1.6 \times 10^{-3}$ & 0.7
&$9.1\times 10^{-4} $& $-3.8\times 10^{-2}$ & $-2.75\pm 0.019$ &4.6 &-0.17\\
RH5 &  $4\times10^{-4}$ & $2.9\times 10^{-3}$ & $2.4\times 10^{-2}$ & $9\times 10^{-2}$ & $4.4\times 10^{-3} $&$-7.6 \times 10^{-2} $& $-5.2\pm0.2$&0.7&2.04\\
RH10 & $1\times10^{-3}$  & $3.6\times 10^{-3}$ & $1.3\times 10^{-2}$  & $7.3 \times 10^{-2}$  &$1.9 \times 10^{-3}$  &$ -9.6\times10^{-2}$ & $-7.2\pm0.7$&0.3&4.2\\
RH20 &  $3\times10^{-3}$ & $5.6\times 10^{-3}$ & $3.9\times 10^{-2}$ &0.35  & $-9.7\times 10^{-3}$ &$3.5 \times 10^{-2} $& $2.44\pm0.08$&1.9&1.3\\
RH40  & $5\times10^{-3}$  & $4.7\times 10^{-3}$ & $0.14$ & 0.35 & $2.9 \times 10^{-2}$  &$-2.5\times 10^{-2} $& $-1.5\pm0.24$&2.3&1\\
RH100&$9\times10^{-3}$ &  $3.6\times 10^{-3}$ & $2.2\times 10^{-2}$  & 0.33 &$  -5.1 \times 10^{-2} $ & $-6.6 \times 10^{-2}$ & $-1.85\pm2.9$&1&1.1\\
\hline
\end{tabular}
\end{table*}

Table~\ref{tab:errors} gathers results of our convergence study.  We compute
relative differences between image integrated $I_{\rm tot}$, 
$LP_{\rm tot} \equiv \sqrt{Q_{\rm tot}^2 + U_{\rm tot}^2}$,
$V_{\rm tot}$, and $\chi_{\rm tot}\equiv arg(Q_{\rm tot}+iU_{\rm tot})/2$ computed for two runs with
different number of radiative transfer integration points along the
geodesics. We find that all maps of I shown in Figure~\ref{fig:imgall} computed
with fully polarized transfer model are consistent with unpolarized images
presented in M16.
We find $\sim 0.1$ per cent convergence in Stokes I, in all cases. For cases
with very high Faraday optical depth, LP and V are poorly converged. In
some models, e.g., RH 40, the relative difference between 
the image integrated V can be as large as 35 per cent when $V \simeq 1\%$.

\subsection{Faraday rotation measure dependency on $\dot{M}$ and disk temperature}

\begin{figure}
\begin{center}
\includegraphics[width=0.5\textwidth,angle=0]{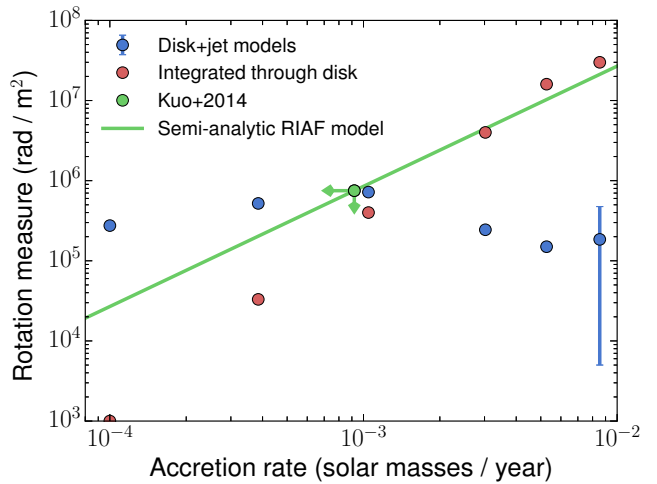}
\caption{Rotation measure as a function of accretion rate onto the black
  hole. The analytically calculated Faraday depth follows a $\tau_{\rm FR} \sim \dot{M}^{3/2} M_{\rm
    BH}^{-1/2}$ relation, based on a model in which the synchrotron source is behind an
  external Faraday screen \citep{marrone:2006}. Data points are from our simulations
  summarized in Table~\ref{tab:errors}. The data point with an arrow is the observed
  upper limit for RM from \citet{kuo:2014}. RMs computed using
  Eq.~\ref{eq:far2} are marked as ``integrated through disk''.}
\label{fig:rmplot}
\end{center}
\end{figure}

We calculate $\chi_{\rm tot}$ which can be used to compute RM as expected from
of non-VLBI observations. The $\chi_{\rm{tot},\nu_1}-\chi_{\rm{tot},\nu_2}$ in
the Table~\ref{tab:errors} is change of total LP angle between 230 and 240
GHz.  Based on that we compute RM. The errorbars of RM is a deviation from
average RM computed based on with models with n=8000, 16000, 32000, and
n=50000 integration points along the geodesics. The errorbars, as expected,
increase with model Faraday depth.  Figure~\ref{fig:rmplot} illustrates the
model RM as a function of mass accretion rate onto the black hole. These are
compared with the \citet{kuo:2014} M87* measurement and the analytic
expectation $\tau_{\rm FR} \sim \dot{M}^{3/2} M_{\rm BH}^{-1/2}$ from external
Faraday rotation in a cold accretion flow. We find that the self-consistently
computed RM first slowly 
increases and then decreases with increasing mass accretion rate and $R_{\rm high}$.
The exact rate of the initial increase of RM 
will depend on details of the electron temperature model.
In the present set up, models with mass accretion rate $> 9 \times
10^{-4} \mdotu$ are consistent with non-VLBI observations of the M87*. 
Faraday RMs based on our fully self-consistent calculations
significantly differ from those computed using Eq.~\ref{eq:far2} and shown in
Figure~\ref{fig:rmplot} as red dots.

\begin{figure*}
\begin{center}
\includegraphics[width=0.6\textwidth,angle=0]{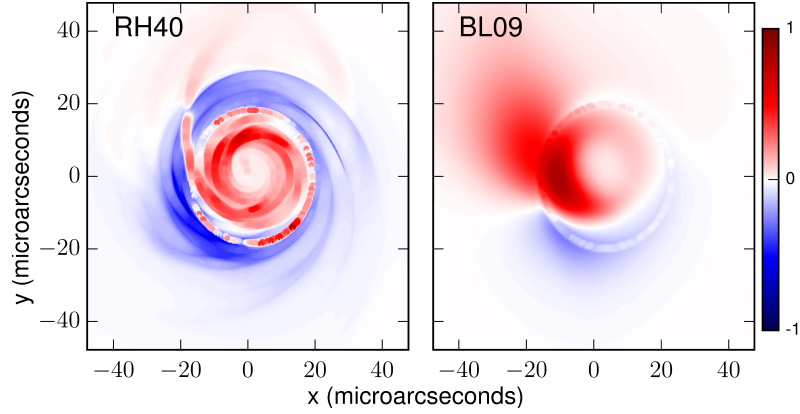}
\caption{Intensity for each image pixel originating from
  below (blue) and above (red) the midplane for the RH40 (left) and
  BL09 (right) models. Each image is scaled linearly to its maximum intensity.
  The forward jet, above the midplane, produces
  most of the observed emission in the BL09 case. In the RH40 case,
  the bulk of the flux density originates in the counter-jet, with a
  contribution from the forward jet.}
\label{fig:midplane}
\end{center}
\end{figure*}

\begin{figure*}
\begin{center}
\includegraphics[width=0.6\textwidth,angle=0]{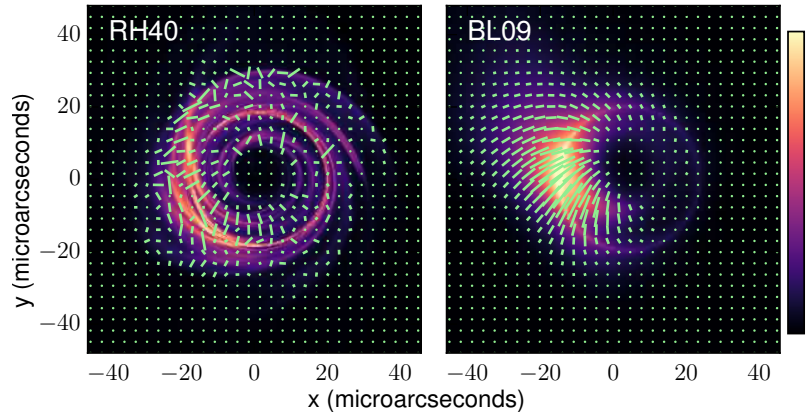}
\caption{Intensity (colors) and polarization maps (ticks) for model
  RH40 (left) and a semi-analytic
  force-free jet model \citep[right,][BL09]{broderickloeb2009}. 
Each image is scaled linearly to its maximum intensity.
The strong
Faraday rotation through the accretion disk leads to a scrambled
polarization pattern in the RH40 case, while the force-free jet shows
coherent polarization which traces its helical magnetic field
structure. The BL09 model has much higher net polarization ($\simeq
15$ per cent compared to $\simeq 1$ per cent for RH40), which is not seen in SMA
observations of M87* \citep{kuo:2014}.
}
\label{fig:imgbl09}
\end{center}
\end{figure*}

Moving from the disk emission dominated models (RH1) to jet emission dominated models
(RH40, RH100) the accretion rate increases but at the same time the electron
temperature in the disk decreases, which shifts the emission region closer towards the
jet where the electrons remain relativistically hot. 
In model RH40, the millimetre emission (Stokes I) is predominantly produced by the jet
sheath. The emission from the jet sheath is
produced just near the event horizon of the black
hole at the jet launching point. For a viewing angle of 20 degrees an external
observer sees a ring of emission around the shadow of the black hole that is a
gravitationally lensed image of the counter-jet. Weak emission from the
forward jet can also be noticed. This is illustrated in Figure~\ref{fig:midplane}. 
The counter jet dominates the emission
because it arises at a small projected separation from the black hole, where the light
bending effects are strong and the beaming comes from the orbital rather
than radial velocity \citep{dexteretal2012}. 
For comparison we also plot an image from the force-free jet
model of \citet{broderickloeb2009}. The model parameters are the same
as for their M0 model, but with a footpoint radius of $r_{\rm fp} =
2$M, for better agreement with the compact observed 230 GHz size of
M87* \citep{doeleman:2012}. 
 Figure~\ref{fig:imgbl09}, shows the significant difference in
  polarization structure between both models. 
Spatially resolved polarization observations could distinguish
between the \citet{broderickloeb2009} forward jet and the RH40
counter-jet models.

\begin{figure*}
\begin{center}
\includegraphics[width=0.3\textwidth,angle=-90]{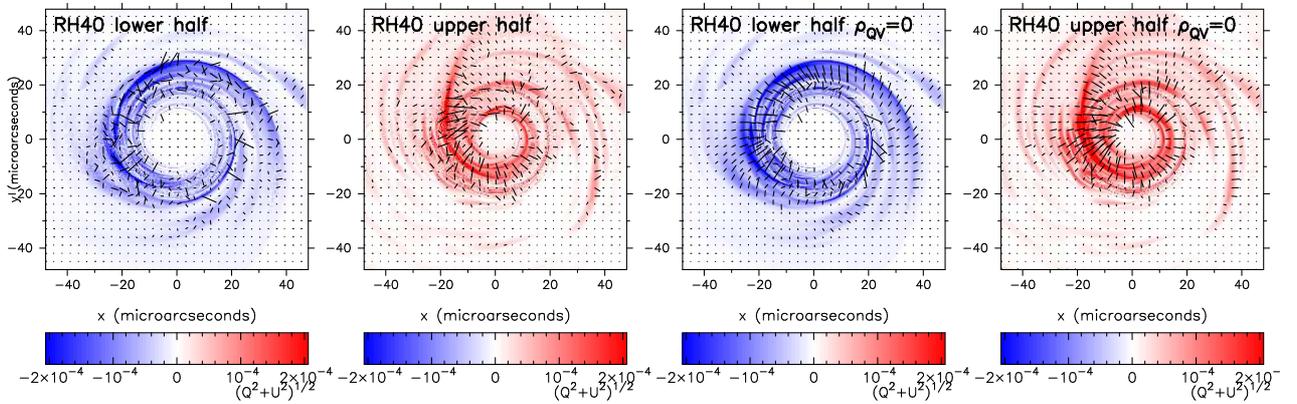}
\caption{First and second panels: polarized emission $\sqrt{Q^2+U^2}$ for each pixel originating from
below (blue) and above (red) the midplane together with polarization ticks 
for model RH40. Third and fourth panels: same as first and second panels but with Faraday effects switched off.
The model without Faraday effects shows coherent polarization signals from both counter- and
forward-jet.}
\label{fig:dep}
\end{center}
\end{figure*}

Synchrotron radiation is intrinsically highly polarized \citep[up to 100 per cent, e.g.,][]{pandya:2016}, but in
our images the brightest regions (in Stokes I) are depolarized to 10 per cent and the total
linear polarization degree is 1 per cent on average (in all models except in model RH1, see the
last column in Table~\ref{tab:errors}). What is causing the strong depolarization?  
We recompute model RH40 using lower and upper hemispheres of the GRMHD
simulations to investigate the polarimetric properties of light coming from
the counter-jet and forward-jet, respectively. 
Fig.~\ref{fig:dep} shows the structure of linear polarization intensity and orientation of the polarization
plane of the counter- and forward-jet.
In Fig.~\ref{fig:dep}, the counter jet polarization (first
panel) is evidently significantly scrambled compared to coherent signal from
the forward jet (second panel). 
The total LP degree from the counter-jet is 1 per cent. This is smaller than
the total polarization degree of the forward-jet which is 3.1 per cent.
The forward to counter jet LP flux ratio is about 2.2. 
In Fig.~\ref{fig:dep} third and fourth panels, both models are then recomputed
assuming zero Faraday rotation and 
conversion $\rho_{\rm QV}=0$ in Eq.~\ref{eq:polartrans}.
Notice that if $\rho_{\rm QV}=0$
the polarization ticks are organized in both counter- and forward-
jet. Without Faraday effects, the forward to counter jet LP flux ratio is
1.1 indicating that indeed the Faraday effects depolarize the counter jet emission
and scramble its polarization structure. 
Here, the total LP degree is 3 and 6 percent for counter and forward-jet
respectively. Total low polarization degree in models with $\rho_{\rm QV}=0$ are 
due to a varying magnetic field structure across the image (``beam depolarization'').

\begin{figure*}
\begin{center}
\includegraphics[width=0.37\textwidth,angle=-90]{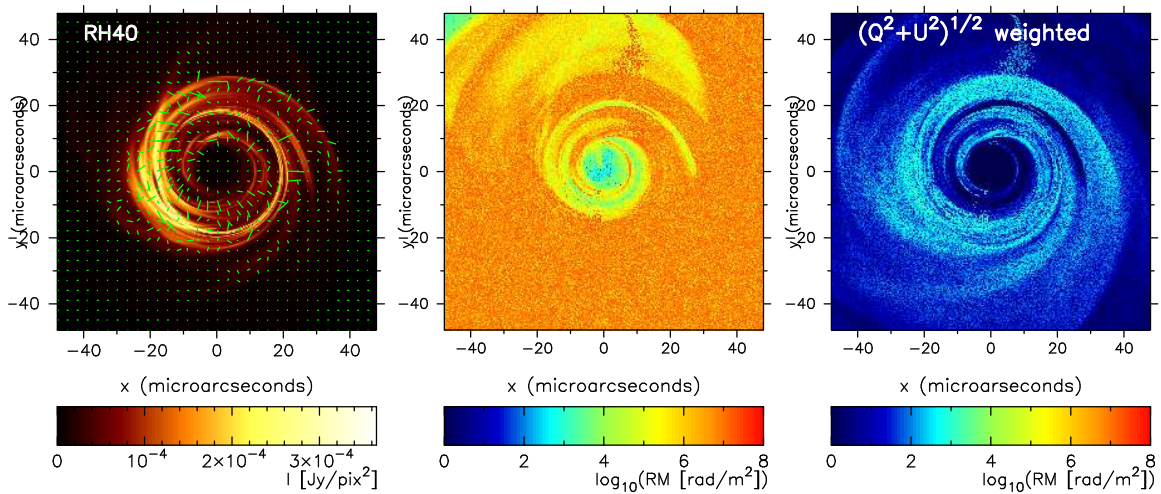}
\caption{Left panel: Intensity map of model RH40 at 230 GHz overplotted with
  linear polarization $\chi$ ticks. Middle panel: RM computed from maps as in the left panel at two frequencies 230 and 240 GHz.
  The right panel: the same as middle panel but RM map is weighed with linear
  polarization $\sqrt{Q^2+U^2}$ to demonstrate how the observed linearly polarized
  emission is affected by Faraday rotation.}
\label{fig:RMmap}
\end{center}
\end{figure*}

If the majority of the observed LP in model RH40 
is produced by the forward jet (because the counter-jet is
depolarized) then the observed RM is probing the plasma conditions in the
forward-jet. The effect is illustrated in Figure~\ref{fig:RMmap}.
The left panel in Figure~\ref{fig:RMmap}, shows the total intensity maps
(orange) with linear polarization ticks at 230 GHz. The middle panel in
Figure~\ref{fig:RMmap} show maps of RMs computed using Eq.~\ref{eq:far1}.  The
values of RM in the middle panel are up to about $10^7 \, {\rm rad \, m^{-2}}$. 
In  Figure~\ref{fig:RMmap} (right panel), we show the RM map weighted with LP to show how
the observed linearly polarized emission is affected by Faraday rotation
effects. This indicates that the polarized emission
that is supposed to be strongly Faraday rotated is
depolarized, and we pick up signals only from the polarized emission in front of the
accretion disk that experiences much weaker Faraday rotation. As a consequence the
measured RM is fairly independent of the disk density ($\dot{M}$) or temperature ($R_{\rm high}$).

\begin{figure*}
\begin{center}
\includegraphics[width=0.9\textwidth,angle=0]{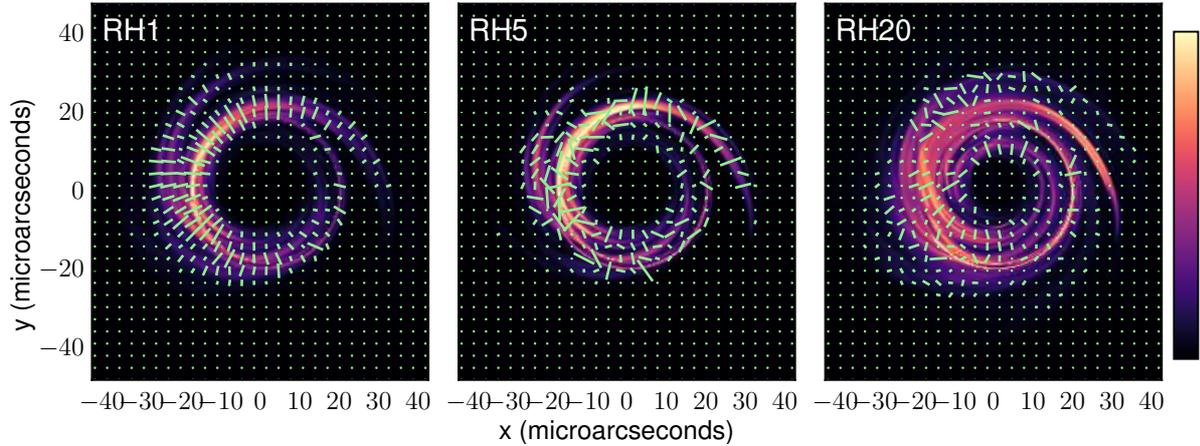}
\caption{Intensity maps of model RH1-20 at 230 GHz overplotted with
  linear polarization $\chi$ ticks.  Each image is scaled linearly to its maximum intensity.
The polarization pattern in the RH1 model varies smoothly over the image. The other models have
  lower disk electron temperatures and higher accretion rates. The
  resulting large Faraday optical depth through the disk scatters the
  polarization for lines of sight to the bright, lensed counter-jet.}
\label{fig:imgall}
\end{center}
\end{figure*}

\begin{figure*}
\begin{center}
\includegraphics[width=0.9\textwidth,angle=0]{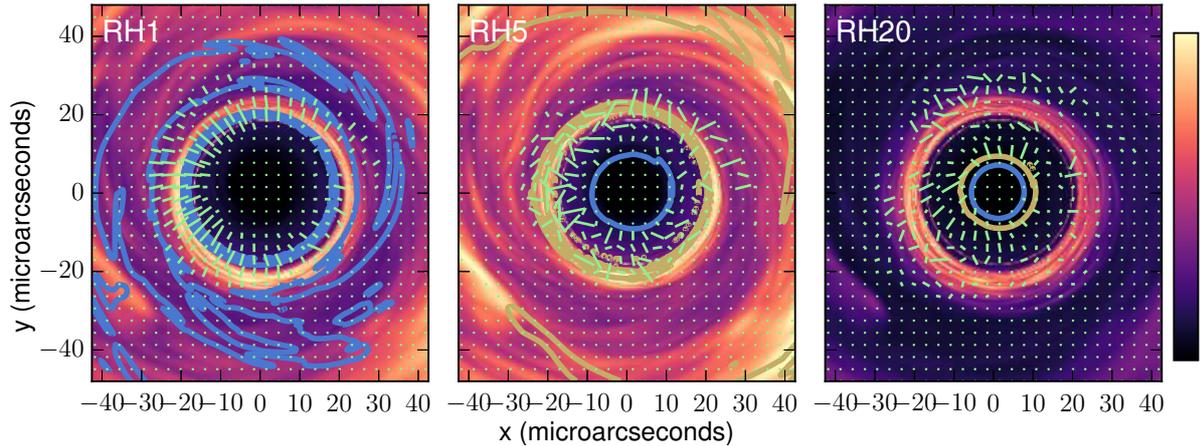}
\caption{Faraday rotation maps (colors) with polarization ticks.
 The blue and yellow contours are $\tau_{\rm FR}$ = 1 and 100,
 respectively. In RH1 many lines of sight are Faraday thin, leading to
 a coherent polarization map tracing the underlying magnetic field
 configuration, while in
 RH20-100 every line of sight of interest has $\tau_{\rm FR} \gg 100$,
 leading to a scrambled polarization map.}
\label{fig:RMall}
\end{center}
\end{figure*}

Which other models RH1-RH100 are, similar to RH40, dominated by Faraday rotation
effects and in which models is the RM due to the forward jet? 
Figs.~\ref{fig:imgall} and~\ref{fig:RMall} show how polarization structure
and Faraday depth change when increasing the accretion rate and decreasing the
temperature of the plasma in the accretion disk (going from model RH1 to RH100). 
In Figure~\ref{fig:imgall}, the LP gets scrambled staring from model RH5 for
which $\tau_{\rm FR}>10$.
Figure~\ref{fig:RMall} shows the absolute value of the
intensity-weighted Faraday optical depth along light rays (colors, with blue and
yellow contours at $\tau_{\rm FR} = 1, 100)$ and polarization maps for the same
models. We find that the polarization gets scrambled for lines of sight
outside the photon ring (bright rings of large Faraday depth). In the RH20
model, every line of sight of interest has a Faraday depth > 100.
We conclude that in models RH20 through RH100, the depolarization is
dominated by Faraday rotation that is occurring in the disk, while the
observed polarization and RM are produced in the forward jet and are
therefore independent of the accretion disk model parameters. The
possibility that the RM originates in the jet rather than the extended
accretion flow is natural given the low inclination of M87 jet, and was
considered previously \citep{kuo:2014,li:2016,feng:2016}. We have
demonstrated this with self-consistent MHD disk+jet models that can match
the radio observations.

\subsection{Constraining the inclination angle of M87 jet}

In these models of M87*, the Faraday rotation measure through the cold,
dense accretion disk is large (e.g., Figure~\ref{fig:tau}). At the low
inclination angles $i \simeq 10-25^\circ$ found from superluminal
motions in the
optical jet \citep[e.g.,][]{biretta:1999}, the line
of sight does not pass through the accretion disk, so that the RM in
our models is produced instead in the forward jet (Figure~\ref{fig:dep}) and does not vary as expected with black hole accretion
rate (Figure~\ref{fig:rmplot}). For larger viewing angles, the line of
sight does pass through the accretion disk.

\begin{figure}
\begin{center}
\includegraphics[width=0.5\textwidth,angle=0]{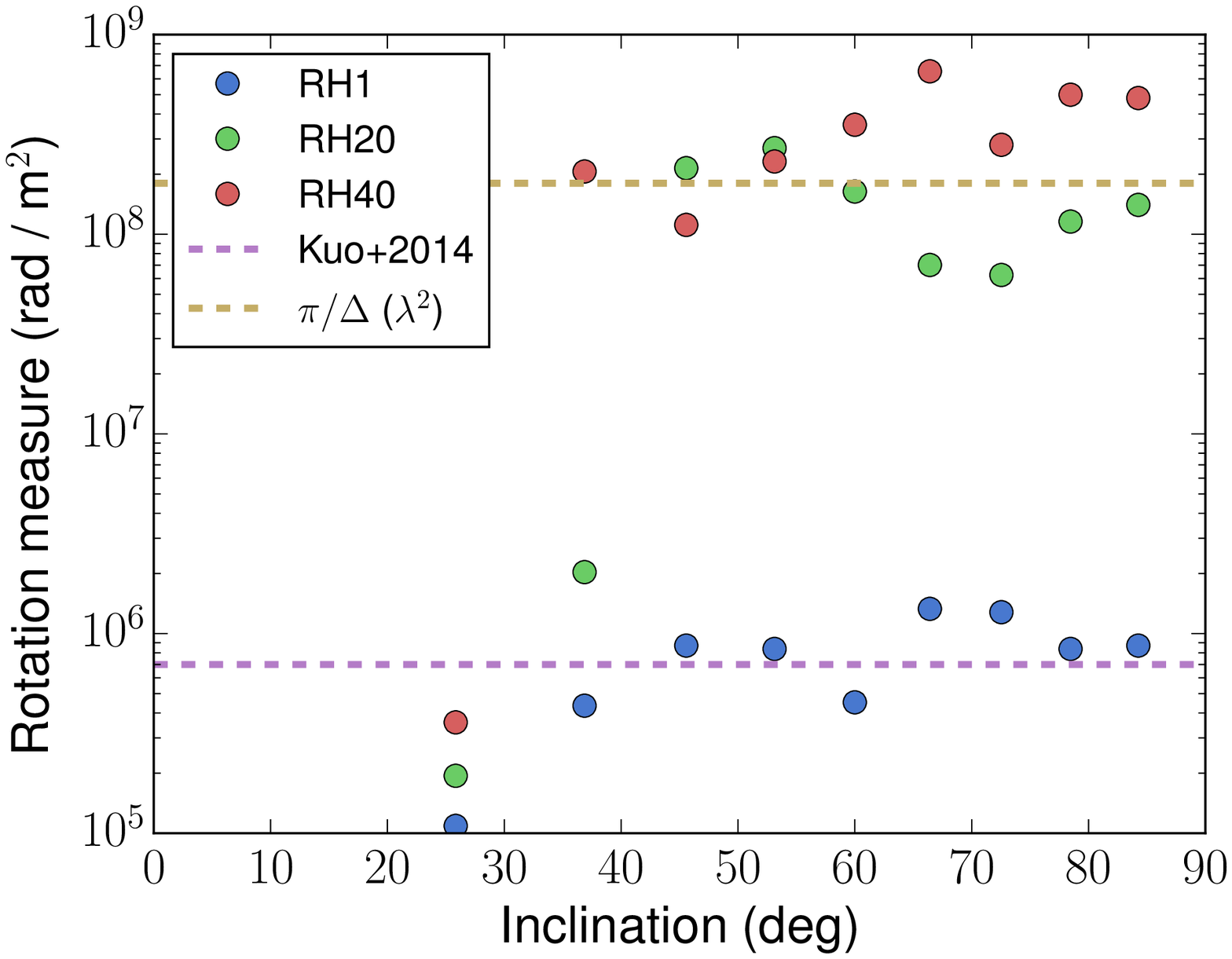}
\caption{Faraday rotation measure as a function of viewing angle for
  disk (RH1) and jet (RH20, RH40) models. In all models, the RM
  increases with increasing inclination due to a growing accretion
  disk contribution. In the jet models, the disk is cold and dense,
  leading to RMs much larger than observed for $i \gtrsim
  30^\circ$. In these cases, the RM quickly saturates near the maximum
  value we can resolve from our calculations with frequency spacing $\Delta \nu = 1$ GHz
centered on $230$ GHz, $\pi / (\lambda_1^2 - \lambda_2^2) \simeq 2
\times 10^8 \hspace{2pt} \rm rad \hspace{2pt} \rm m^{-2}$.}
\label{fig:RMinc}
\end{center}
\end{figure}

We calculate the RM as a function of inclination angle by measuring
$\chi(\nu)$ from the full radiative transfer results at 20 frequencies
between $220-240$ GHz, spaced by $\Delta \nu = 1$ GHz. We fit a model
to the result for the full RM expression,

\begin{equation}
\tan 2\chi = \frac{\tan 2 \chi_0 \cos{RM \lambda^2} - \sin{RM
    \lambda^2}}{\tan 2 \chi_0 \sin{RM \lambda^2} + \cos{RM \lambda^2}}
\end{equation}

\noindent where $\lambda$ is measured in meters and $\chi_0$ and $RM$
are the fit parameters. The result is shown in
Figure~\ref{fig:RMinc}. The RM increases rapidly for inclination angles $ i \gtrsim
30^\circ$ as the line of sight passes through the disk, and saturates
near the maximum value we can measure given our frequency spacing,
$RM_{\rm max} \simeq \pi / \Delta (\lambda^2) \simeq 2 \times 10^8
\hspace{2pt} \rm
rad / \hspace{2pt} \rm m^2$. Models with viewing angle $i>30^\circ$
greatly overproduce the observed RM \citep[purple dashed
line,][]{kuo:2014}. In this way, the measured RM provides an
independent constraint on the inclination angle to the M87 jet. The models which best match observations
of the radio jet on larger scales (RH40, RH100) provide the strongest
constraints on the viewing angle, placing a limit of $i \lesssim 30^\circ$.

\section{Summary}\label{sec:dis}

Disk+jet models of the M87* based on GRMHD simulations
\citep{moscibrodzka:2016} produce submillimetre images which are dominated
either by the disk (low $\dot{M}$, hot disk electrons) or the
\emph{counter-jet} (high $\dot{M}$, cold disk electrons). The jet dominated
models are consistent with the observed radio spectrum and jet
morphologies. At short wavelengths the counter-jet produces the emission due
to strong light bending effects (as first shown in \citealt{dexteretal2012}). The cold accretion
disk in these models Faraday depolarizes the counter-jet.

We find that in the majority of models Faraday rotation in the cold accretion disk
depolarizes the counter-jet emission. The small residual polarization comes
mostly from the forward jet and does not pass through the accretion disk, so
that the observed Faraday rotation measure (RM) may be smaller than
anticipated for a given mass accretion rate and using simple models.
All models considered, even
with accretion rates a factor $\simeq 10$ larger than the limit from
\citet{kuo:2014}, are still consistent with the measured RM of the M87*. 
It is possible that viable models could have even higher accretion rates
than considered here ($\dot{M}>9\times10^{-3}\, \mdotu$).

It is difficult to integrate the equations of polarized transfer with high
accuracy in models with high Faraday depth.  The inaccuracy is probably
due to spacing of the geodesic points concentrated close to the black hole
event horizon in the current scheme (and in most of similar types of
codes). While this is convenient to produce accurate intensity images a lot of
Faraday rotation (or conversion) can occur far from the black hole where the
integration is not spaced well. This should be addressed by future
radiative transfer models.

We examined a specific type of GRMHD simulations (so called standard
and normal evolution models, SANEs). There are other possible models
for the M87* (e.g. Magnetically Arrested Disks,
\citealt{mckinney:2012}), which we have not explored further.  It
should be feasible to check how RM depends on, e.g., mass accretion
rate in these other models to find out if they are consistent with
observations. In any case, our models predict different levels of
polarization degree and possibly different Faraday rotation compared
to semi-analytical models of the jet launching zone as presented in
\citet{broderickloeb2009}.

What are the implications of our results for other low luminosity supermassive
black hole systems? Based on variability timescales analysis
\citet{bower:2015} and spectral modelling \citet{falcke:1996} 
the millimetre emission from the core of M81 (M81*)
should also be coming from the immediate vicinity of the black hole event
horizon (i.e., $\tau_{\rm abs} \leq 1 $ at mm wavelengths).  The Faraday rotation
depth dependency on the mass of the black hole $M_{\rm BH}$ is $\tau_{\rm FR}
\propto \dot{M}^{3/2} M_{\rm BH}^{-2} R_{\rm high}^2 $ (because $n_e \propto
\dot{M}/M_{\rm BH}^2$ and $B \propto \dot{M}^{1/2}/M_{\rm BH}$ but $dl \propto
M_{\rm BH}$).  Combining the above expression with numerical values of
$\tau_{\rm FR}$ presented in Figure~\ref{fig:tau} we obtain an approximate general
expression for the Faraday depth as a function of black hole mass, accretion
rate, and proton-to-electron temperature ratio in the disk:
\begin{equation}
\tau_{\rm FR} \approx 1 (\frac{\dot{M}}{10^{-4} \mdotu})^{3/2} (\frac{M_{\rm
    BH}}{6.2 \times 10^9 \msun})^{-2} (R_{\rm high})^2. 
\end{equation}
Assuming that the mass of the supermassive black hole in M81* is $M_{\rm BH}=7\times10^7
M_\odot$ and, e.g., $R_{\rm high}=10$ we find that $\tau_{\rm FR} \gg 1 $ for
accretion rates $\dot{M}> 10^{-7} \,\mdotu$. Hence, we conclude that it is likely
that the effects presented in this work also will be visible in M81*,
i.e., the millimetre source polarization will be scrambled and depolarized.

\section*{Acknowledgements}
M. Moscibrodzka and J.Dexter contributed equally to this work.
M. Moscibrodzka, J. Davelaar and H. Falcke acknowledge support from the ERC
Synergy Grant "BlackHoleCam-Imaging the Event Horizon of Black Holes" 
(Grant 610058). J. Dexter was supported by a Sofja
Kovalevskaja Award from the Alexander von Humboldt Foundation of
Germany. The authors thank BlackHoleCam collaborators, C.F. Gammie and
C.v. Eck 
for comments on the manuscript.




\bibliographystyle{mnras}
\bibliography{local} 



\appendix
\onecolumn
\section{Polarized radiative transfer in spherical coordinates}
\label{sphstokes}

The non-relativistic polarised radiative transfer equation can be written in the form,

\begin{equation}\label{eq:polartrans}
\frac{d}{ds}
\left(\begin{array}{c}  I \\  Q \\  U \\  V \\\end{array}\right)
=
\left(\begin{array}{c}
  j_I \\  j_Q \\  j_U \\  j_V
\end{array}\right)-
\left(%
\begin{array}{cccc}
  \alpha_I & \alpha_Q & \alpha_U & \alpha_V \\
  \alpha_Q & \alpha_I & \rho_V & \rho_U \\
  \alpha_U & -\rho_V & \alpha_I & \rho_Q \\
  \alpha_V & -\rho_U & -\rho_Q & \alpha_I \\
\end{array}%
\right)
\left(\begin{array}{c}  I \\  Q \\  U \\  V \\\end{array}\right)
\end{equation}

\noindent where ($I$, $Q$, $U$,
$V$) are the Stokes parameters, $j_{I,Q,U,V}$ are the polarised
emissivities, $\alpha_{I,Q,U,V}$ are the absorption coefficients, and
$\rho_{Q,U,V}$ are the Faraday rotation and conversion coefficients. 

In the absence of emission and absorption, the Faraday rotation
coefficient $\rho_V$ produces oscillations in $Q$ and $U$,
corresponding to a rotation of the polarization angle. As described in
the text, numerical solutions to the equations as written above break
down when the oscillation length is shorter than the radiative
transfer grid spacing $\Delta s$. In this limit, we follow
\citet{romans:2012} and transform the polarized Stokes parameters to
``spherical'' coordinates:

\begin{equation}\label{transfer}
\left(\begin{array}{c}  Q \\  U \\  V \\\end{array}\right)
=
\left(\begin{array}{c}
    R_S \sin{\psi_S} \cos{\phi_S} \\ R_S \sin{\psi_S} \sin{\phi_S} \\ R_S
        \cos{\psi_S}\end{array}\right),
\end{equation}

\noindent where $R_S^2 = Q^2+U^2+V^2$ is the polarized intensity and
$\phi_S$ and $\psi_S$ are angular Stokes parameters describing the
linear polarization direction ($\phi_S$) and the relative amount of
circular polarization ($\psi_S$). In terms of these spherical
Stokes parameters, the transfer equations become:

\begin{eqnarray}
\frac{dR_S}{ds} &=& \sin{\psi_S}\left[\cos{\phi_S}\left(j_Q-\alpha_Q
  I\right)+\sin{\phi_S}\left(j_U-\alpha_U
  I\right)\right]+\cos{\psi_S}\left(j_V-\alpha_V I\right)-\alpha_I
                    R_S,\\
\frac{d\phi_S}{ds} &=& \frac{\csc{\psi_S}}{R_S}
                       \left[\cos{\phi_S}\left(j_U-\alpha_U
                       I\right)-\sin{\phi_S}\left(j_Q-\alpha_Q
                       I\right)\right]-\cot{\psi_S}\left(\cos{\phi_S}
                       \rho_Q + \sin{\phi_S} \rho_U\right)+\rho_V,\\
\frac{d\psi_S}{ds} &=& \frac{1}{R_S} \left\{\cos{\psi_S}
                       \left[\cos{\phi_S}\left(j_Q-\alpha_Q
                       I\right)+\sin{\phi_S}\left(j_U-\alpha_U
                       I\right)\right]-\sin{\psi_S}\left(j_V-\alpha_V
                       I\right)\right\}+\cos{\phi_S} \rho_U-\sin{\phi_S}\rho_Q.
\end{eqnarray}

\noindent In this form, the transfer equations are non-linear
and must be solved by numerical integration. On the other hand, the
Faraday effects now only appear in equations for the angular Stokes
parameters, and so do not directly change the total polarized intensity,
$R_S$. Further, these effects are now source terms, independent of the
spherical Stokes parameters. When they change rapidly (e.g. $\Delta \phi_S
\gg 2\pi$) during a single radiative
transfer step, the accuracy in the solutions for the polarization
angle and relative linear/circular polarization decreases. The total
intensity and polarization fraction remain unchanged, however,
allowing the integration to proceed even if the resulting 
Stokes $Q$, $U$, and $V$ are noisy (Table~\ref{tab:errors}).


\bsp	
\label{lastpage}
\end{document}